\documentclass[aip, amsmath,amssymb, reprint, superscriptaddress]{revtex4-2}

\usepackage{graphicx}
\newcommand{\RNum}[1]{\uppercase\expandafter{\romannumeral #1\relax}}
\usepackage{amsmath}
\usepackage{amssymb}
\usepackage{bm}
\usepackage{float}
\usepackage{dcolumn}
\usepackage{subfigure}
\usepackage{units}
\usepackage{hyperref}
\usepackage[usenames]{color}
\usepackage[dvipsnames]{xcolor}
\usepackage{booktabs}
\usepackage{multirow}
\usepackage{wrapfig}
\usepackage{lipsum}
\usepackage{setspace}
\usepackage{xcolor}
\usepackage{longtable}
\hypersetup{pdfborder=0 0 0,colorlinks=true,citecolor=blue,linkcolor=blue}

\usepackage{gensymb}

\usepackage{ulem}

\begin{document}

\preprint{AIP/123-QED}

\title[]{Antiferromagnetic droplet soliton driven by spin current}
\author{Roman V. Ovcharov}
\affiliation{ 
Department of Physics, University of Gothenburg, Gothenburg 41296, Sweden
}

\author{Mohammad Hamdi}
\affiliation{Ecole Polytechnique Federale de Lausanne (EPFL), Institute of Materials,
Laboratory of Nanoscale Magnetic Materials and Magnonics, CH-1015 Lausanne, Switzerland}
\affiliation{Department of Electrical and Computer Engineering, Northwestern University, Evanston, IL 60208, USA\looseness=-1}

\author{Boris A. Ivanov}
\affiliation{
Institute of Magnetism of NASU and MESU, Kyiv 03142, Ukraine
}
\affiliation{
Radboud University, Institute for Molecules and Materials, Nijmegen 6525 AJ, Netherlands\looseness=-1
}

\author{Johan \AA kerman}
\affiliation{ 
Department of Physics, University of Gothenburg, Gothenburg 41296, Sweden
}
\affiliation{
Center for Science and Innovation in Spintronics, Tohoku University, Sendai 980-8577, Japan\looseness=-1
}
\affiliation{
Research Institute of Electrical Communication, Tohoku University, Sendai 980-8577, Japan\looseness=-1
}

\author{Roman S.  Khymyn}%
\affiliation{ 
Department of Physics, University of Gothenburg, Gothenburg 41296, Sweden
}
\email{roman.khymyn@physics.gu.se}
\date{\today}

\begin{abstract}
We demonstrate that a spin current flowing through a nano-contact into a uniaxial antiferromagnet with first- and second-order anisotropy can excite a self-localized dynamic magnetic soliton, known as a spin-wave droplet in ferromagnets. 
The droplet nucleates at a certain threshold current with the frequency of the N\'eel vector precession laying below the antiferromagnetic resonance. The frequency exhibits nonlinear behavior with the increasing of applied current. At the high value of applied torque, the soliton mode transforms, and the oscillator emits spin waves propagating in the antiferromagnetic layer.
\end{abstract}

\maketitle

Antiferromagnetic materials (AFMs) have unique properties advantageous for future spintronic applications, including the absence of stray fields, intrinsic high-frequency dynamics, high spin wave velocities, and abundance in nature \cite{Zhang2020,Austefjord2022}. Utilizing their terahertz (THz) spin dynamics due to strong exchange interaction can bring about solid-state THz nano-devices and, hence, close the THz gap \cite{Sirtori2002,Osborne2008}. One of the most promising candidates of such devices is the AFM-based spin-Hall and spin-transfer torque nano-oscillators (SH/ST-NOs), which can operate as THz sources and detectors \cite{Cheng2016,Khymyn2017a,Sulymenko2017,Johansen2017,Khymyn2017,Gomonay2018,Puliafito2019,Lee2019,Lisenkov2019,Troncoso2019,Parthasarathy2021,Sulymenko2018,Khymyn2018}. Furthermore, these devices offer great potential for on-chip THz neuromorphic applications \cite{Sulymenko2018,Khymyn2018,Sulymenko2019,Sulymenko2019a,Grollier2020a}. Several attempts have been made to understand current-driven spin dynamics in single SH/ST-NOs \cite{Cheng2016,Khymyn2017a,Sulymenko2017,Johansen2017,Khymyn2017,Gomonay2018,Puliafito2019,Lee2019,Lisenkov2019,Troncoso2019,Parthasarathy2021,Sulymenko2018,Khymyn2018}, which can further be coupled by  propagating THz Slonczewski spin waves \cite{Hamdi2022b}, similar to ferromagnetic counterparts\cite{Kendziorczyk2014,Houshang2016,Awad2017,Zahedinejad2020,Zahedinejad2022}. 

Employing \textit{localized} spin dynamics is of crucial importance for the operation of spintronics devices \cite{Gomonay2018a,Chen2023}. Contrary to ferromagnets (FMs), where a combination of demagnetization, crystal anisotropy, and external field can form the localizing potential for magnons, the localization of spin dynamics in AFMs is challenging. It can be achieved by exciting self-localized AFM spin textures such as domain walls, Bloch lines, and skyrmions \cite{Ovcharov2022,Ovcharov2023a, Chen2023}, however pure dynamical localized excitations can substantially enrich the scopes of AFM devices. Such dynamical states in the form of AFM solitons were predicted theoretically a long time ago \cite{Kosevich1990,Baryakhtar1983,Galkina2018} in the case of zero damping. However, their practical realization was unresolved due to the lack of any excitation method. In contrast, FM dynamic solitons, such as droplets, are experimentally demonstrated in SH/ST-NOs \cite{Mohseni2013,Mohseni2018,Ahlberg2022,Chung2016,Divinskiy2017,Macia2020}.
\begin{figure}[t!]
\centering
	\includegraphics[width=0.85\linewidth]{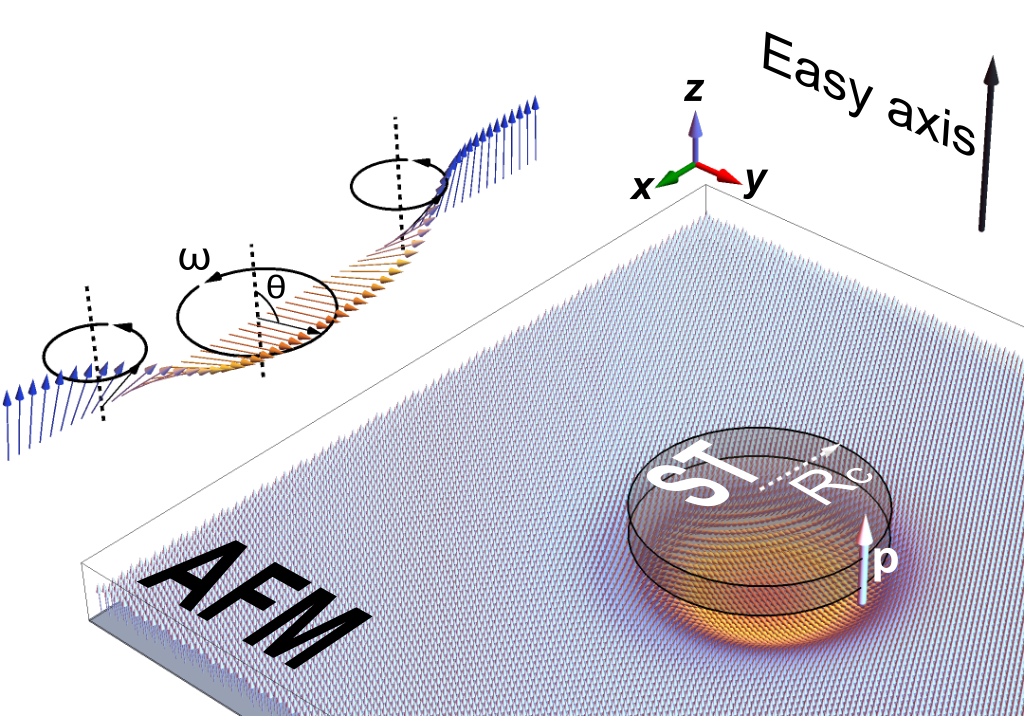}
	\caption{Schematic illustration of an AFM ST oscillator. The nanocontact, which acts as a spin current source, is placed on top of a thin AFM layer with uniaxial anisotropy. The black arrow shows the easy-axis orientation, and the white arrow indicates the direction of the spin current polarization. In the upper left corner, a sketch of the N\'eel vector precession shows the spin structure across the excited droplet. }
	\label{fig:schema}
\end{figure}

In this Letter, we study the excitation of dissipative AFM droplet solitons in a nanocontact (NC)-based SH/ST-NO. We use micromagnetic simulations to investigate the stability and properties of the excited AFM droplets as a function of applied current, magnetic anisotropies, and NC radius. In particular, we compare droplet structures for different NC radii and evaluate their influence on the output signal. Our choice of material is Ru- and Rh-doped hematite ($\alpha$-Fe$_2$O$_3$), which has been identified as a promising candidate for potential experimental realization \cite{Hamdi2023b}. 

\textit{Analytical model.} We consider a scheme that is widely used for the excitations of the droplets in ferromagnets and is shown in Fig. \ref{fig:schema}. It consists of an AFM thin film and adjunct NC that is a source of spin current, providing spin-transfer or spin-orbit torque onto the AFM magnetic sublattices. The AFM has uniaxial anisotropy, and spin current is polarized along the easy axis. In ferromagnets strong enough out-of-plane anisotropy, overcoming the demagnetizing field, creates attractive coupling between magnons, which is necessary for a self-formation of droplet-like solitons \cite{Hoefer2010,Mohseni2013,Mohseni2018}. 
Contrary to ferromagnets, simple quadratic anisotropy (in the form $-K_1 M_z^2$) does not provide nonlinear coupling between magnons in AFMs. It was proposed in Refs. \onlinecite{Kosevich1990,Baryakhtar1983} to employ higher-order terms in the anisotropy energy density as 
\begin{equation}
w_a= -K_1\cos^2 \theta - K_2\cos^4 \theta, \quad K_1,K_2>0
\label{anisotropy}
\end{equation}
to stabilize droplets, where $\theta$ is the angle between easy-axis and N\'eel vector. This anisotropy together with the exchange field $H_{ex}$ defines two characteristic frequencies of the AFM: for the small amplitude precession ($\theta\simeq 0$): $\omega_{\text{AFMR}}=\gamma \sqrt{H_{\text{ex}} (K_1+2K_2)/M_s}$ and the maximum one ($\theta\simeq \pi/2$): $\omega_{\text{SF}}=\gamma \sqrt{H_{\text{ex}} K_1/M_s}$. 

Eq. \ref{anisotropy} is reported to describe magnetic anisotropies in hematite \cite{Morrish1995}. These anisotropy terms can be tuned by doping elements \cite{Hayashi2021} in hematite. For example, it is shown that Ru and Rh (Al and Ga) doping increase (decrease) both $K_1$ and $K_2$ \cite{Hayashi2021}. The exchange field can also be reduced by any doping in hematite \cite{Coey1971,Svab1979}. These properties make hematite an ideal candidate for realizing AFM droplets in experiments. 

The conservative dynamics of the AFM droplet can be described in terms of the angular variables for the N\'eel vector $\theta$ and $\phi$, where $\phi$ is the angle in the hard plane. The soliton solution is $\theta = \theta(r)$ and $\phi = \omega t$, where $\theta(r)$ is governed by the equation:
\begin{equation*}
2r_0^2 \left( \frac{d^2 \theta}{d r^2} + \frac{1}{r} \frac{d \theta}{d r} \right) + \sin 2 \theta\left(\frac{\omega^2 - \omega_{\text{c}}^2}{ \omega_{\text{AFMR}}^2 - \omega_{\text{c}}^2} - \cos 2 \theta\right) = 0.
\end{equation*}
This equation together with the boundary conditions $\theta'(0)=0$, $\theta(\infty)=0$, defines the profile of the soliton at a given frequency $\omega$ (in the range $\omega_\text{c}<\omega<\omega_{\text{AFMR}}$) with $\theta(0)=\theta_0$ at the center of the droplet. Here $\omega_{\text{c}} = \sqrt{(\omega_{\text{AFMR}}^2 + \omega_{\text{SF}}^2)/2}$ is the minimum frequency and $r_0=c/\sqrt{\omega_{\text{AFMR}}^2 - \omega_{\text{c}}^2}$ is a characteristic size of a soliton, $c$ is the maximum speed of magnons defined by the exchange interaction\cite{Kosevich1990}. As we will see below, $r_0$ is an important parameter of the AFM material since it defines the required geometry of the NC for the droplet excition. The frequency of precession $\omega$ is the single variable of the droplet, and its profile can be defined completely at a given $\omega$ and analyzed numerically.

The excitation of a \textit{dissipative} droplet by spin current passing through the NC with the radius $R_{\text{c}}$, Fig. \ref{fig:schema}, requires to account the energy balance between the gain and dissipation across the soliton profile. In the stationary regime of a droplet precession, this condition can be expressed as: 
\begin{equation}
\Gamma_{\text{tot}}=\sigma j \int_{0}^{R_{\text{c}}} \dot{\phi} \sin^2 \theta  r dr -\alpha \int_{0}^{\infty} \dot{\phi}^2 \sin^2 \theta r dr=0,
\label{eq:balance}
\end{equation}
where $\alpha$ is a Gilbert damping constant, $j$ is an electrical current density and $\sigma$ describes ST efficiency. The condition of the Eq. (\ref{eq:balance}) selects the particular frequency and, hence, the profile of a droplet.
\begin{figure}[hbt!]
\centering
	\includegraphics[width=0.78\linewidth]{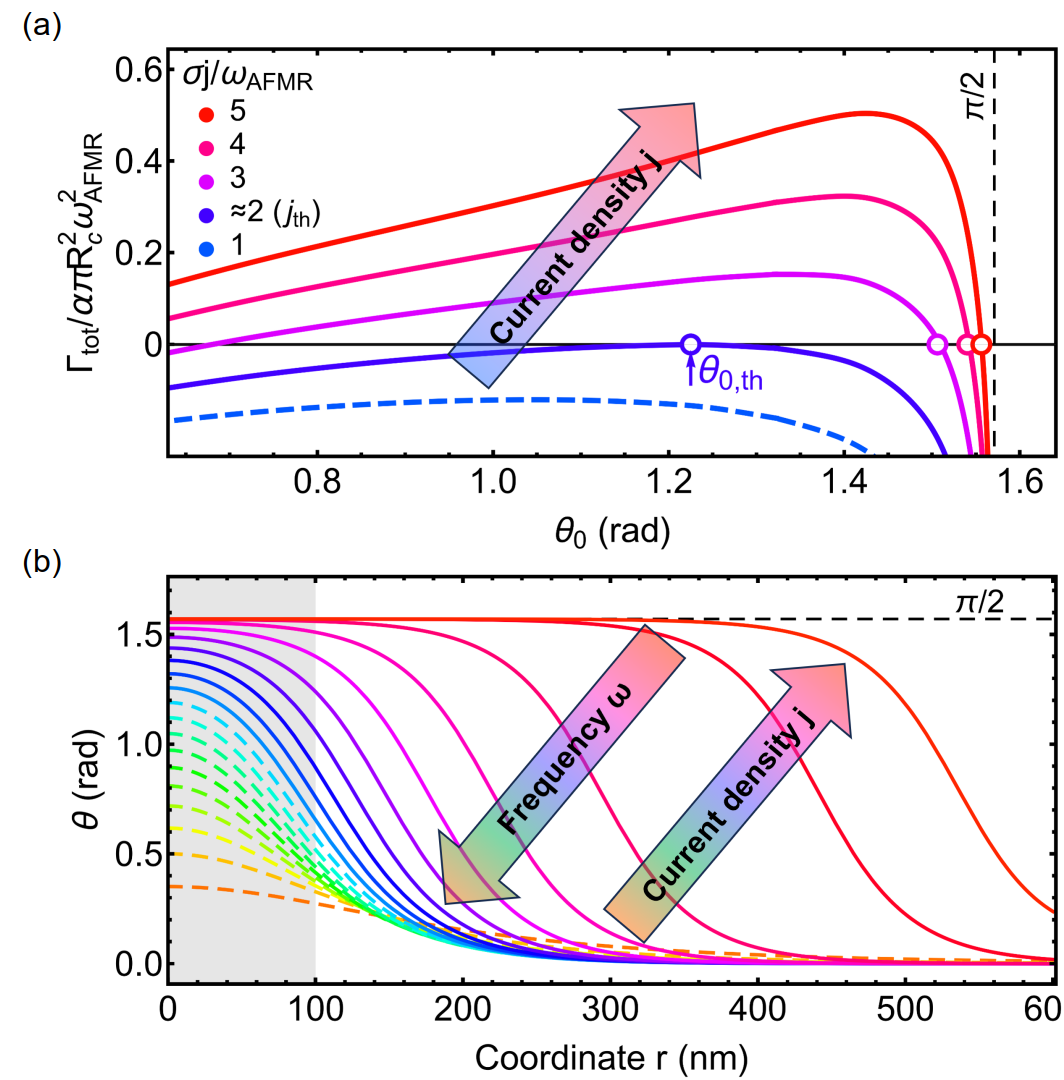}
	\caption{a) Dependence $\Gamma_{\text{tot}}$ on $\theta_0$ for different values of the applied current. The condition $\Gamma_{\text{tot}}=0$ selects the b) profile of the soliton. Profiles with $\theta_0<\theta_{\text{0,th}}$ are shown by dashed lines. Here $\omega_{\text{AFMR}}/2\pi=213$~GHz, $\omega_{\text{c}}/2\pi=192$~GHz, $r_0=40$~nm and $R_{\text{c}}=100$~nm. The direction of the current increase is highlighted by the arrow.}
	\label{fig:analytic}
\end{figure}

The above approach is highlighted in Fig. \ref{fig:analytic}, where $\Gamma_{\text{tot}}$ and droplet profiles are shown at different currents. At the low value of applied current $\Gamma_{\text{tot}}<0$ for all possible $\theta_0$ and the droplet is absent. However at a certain threshold $j_{\text{th}}$ a solution $\Gamma_{\text{tot}}=0$ appears with a finite value of $\theta_0=\theta_{\text{0,th}}$, which in turn correponds to a droplet frequency $\omega_{\text{th}}<\omega_{\text{AFMR}}$. At higher currents, the condition $\Gamma_{\text{tot}}=0$ has two solutions; however, the left one is unstable against an increase in droplet amplitude. While a soliton expands with a current, its frequency gradually decreases towards the limit value $\omega_{\text{c}}$.

\textit{Micromagnetic simulations.} To investigate the dynamics of an AFM droplet, we carried out micro-magnetic simulations using \emph{MuMax3} solver~\cite{Vansteenkiste2014} for a system illustrated in Fig.~\ref{fig:schema}. The structure is composed of an AFM film measuring $516\times516$~nm$^2$ with a thickness of 7~nm. At the center of the device, a circular NC having a radius of $R_{\text{c}}$ is placed, supplying a spin current polarized along the easy axis of the AFM. The AFM material properties are set to correspond to $\alpha$-Fe$_2$O$_3$~\cite{Blake1966,Morrish1995,Turov1983,Turov2001,Hamdi2023,Hamdi2023b,Jani2021}, with sublattice saturation magnetization $M_s=860$~kA/m, exchange stiffness $A_{ex}=7.7$ pJ/m, exchange field $H_{ex}=1800$~T. The intrinsic damping of hematite is rather low  (see, e.g., Ref. \onlinecite{Hamdi2023} with reported $\alpha=1.1\times 10^{-5}$), but to account for the damping enhancement due to the spin pumping, we set it to $\alpha= 10^{-3}$. Similar to the static spin-flop with magnetic anisotropy given by Eq.~\ref{anisotropy}, the excitation induced by the spin current exhibits hysteresis behavior. Consequently, our simulations commence at a higher current, followed by a gradual decrease to the operational value. Hence, the \textit{threshold} current refers to the minimal operational current that maintains excitations. The primary parameters include the excitation frequency $\omega$ under the NC center and the amplitude defined by the deflection angle $\theta_0$ at the same point. To ensure that the excitation corresponds to the droplet mode, we check for the condition for frequency to be lower than AFM resonance $\omega < \omega_{\text{AFMR}}$, where propagating magnons are absent in bulk AFM.

\begin{figure}[hbt!]
\centering
	\includegraphics[width=0.78\linewidth]{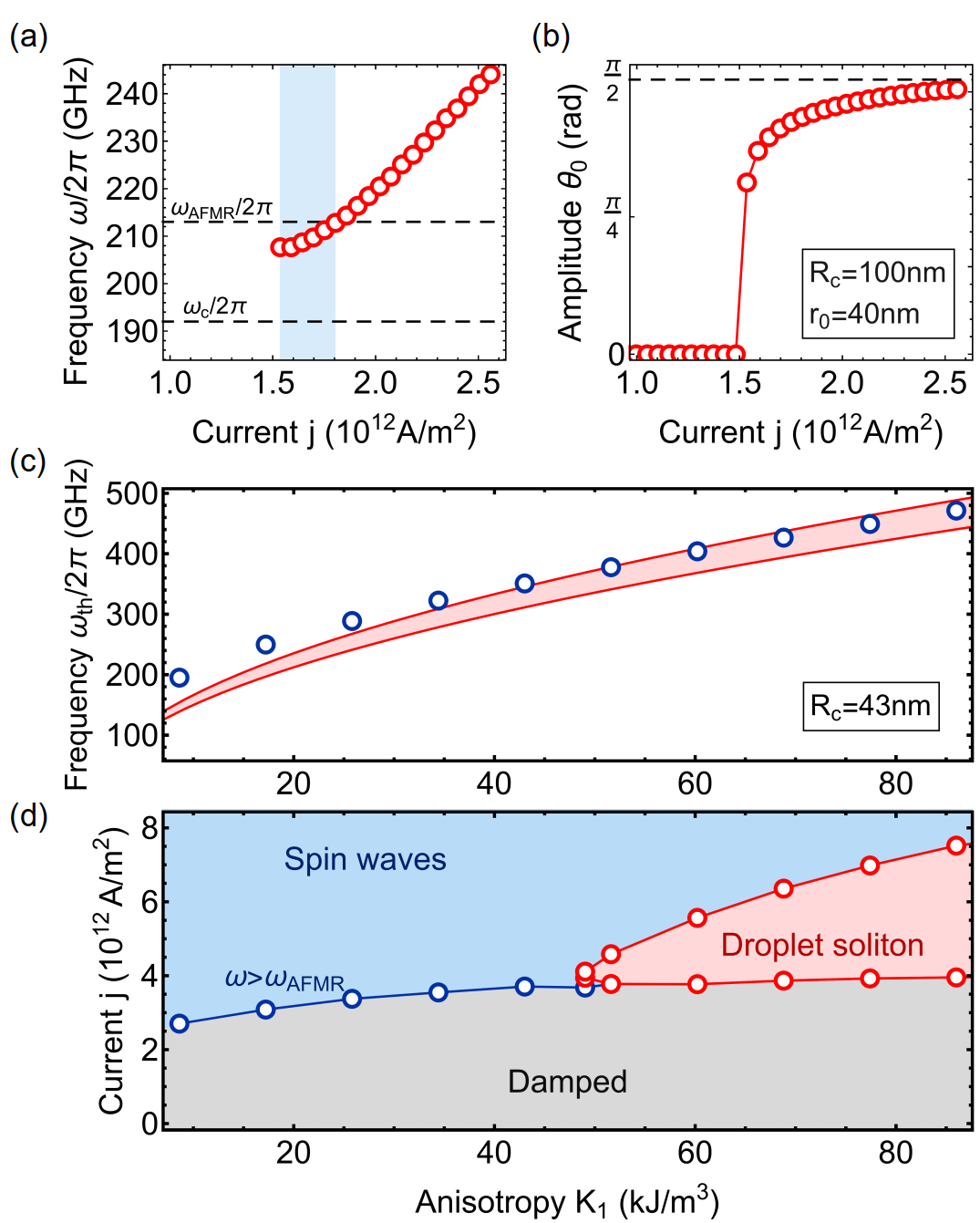}
	\caption{The dependences of (a) the precession frequency $\omega/2\pi$ and (b) the deflection angle $\theta_0$ on the applied current density for the NC with $R_{\text{c}}=100$~nm and characteristic length $r_0 = 40$~nm ($K_1=16.3$~kJ/m$^3$). The dependences on the anisotropy constants of (c) the excitation frequency at the threshold and (d) the ``phase diagram'' of the excitation type as a function of the applied current density. The red-filled region in (c) shows the theoretical limits of the droplet frequency in the non-dissipative limit, indicated by the horizontal dashed lines in (a). The blue bar in (a) highlights the range of currents at which the droplet is observed.}
	\label{fig:anis_results}
\end{figure}

First, we analyzed the case of anisotropy values corresponding to undoped hematite given by $K_1=16.3$~kJ/m$^3$ and $K_2 = 4.9$~kJ/m$^3$, which gives the characteristic length $r_0=40$~nm for the maximum speed of $c = 23$~km/s.
The results of excitation by NC with $R_{\text{c}}=100$~nm are shown in Fig.~\ref{fig:anis_results} (a, b). Notably, a gap between the frequency of the excitation and AFMR appears at the threshold $j_{\text{th}}=1.54\times 10^{12}$~A/m$^2$, similar to what is observed for the droplets in ferromagnetic oscillators. This gap results from an amplitude threshold for the droplet excitation $\theta_0 > \theta_{\text{0,th}}$ discussed above. As the current increases to  $j_{\text{sw}}=1.8\times 10^{12}$~A/m$^2$, the area under the NC starts to emit propagating spin waves instead of the localized droplet. The frequency and applied current range within which the droplet persists is relatively narrow. Furthermore, if the NC radius is reduced to the characteristic size of $r_0$, the initiation of the localized droplet ceases at any current value, yielding only propagating spin waves.

\begin{figure}[hbt!]
\centering
	\includegraphics[width=0.78\linewidth]{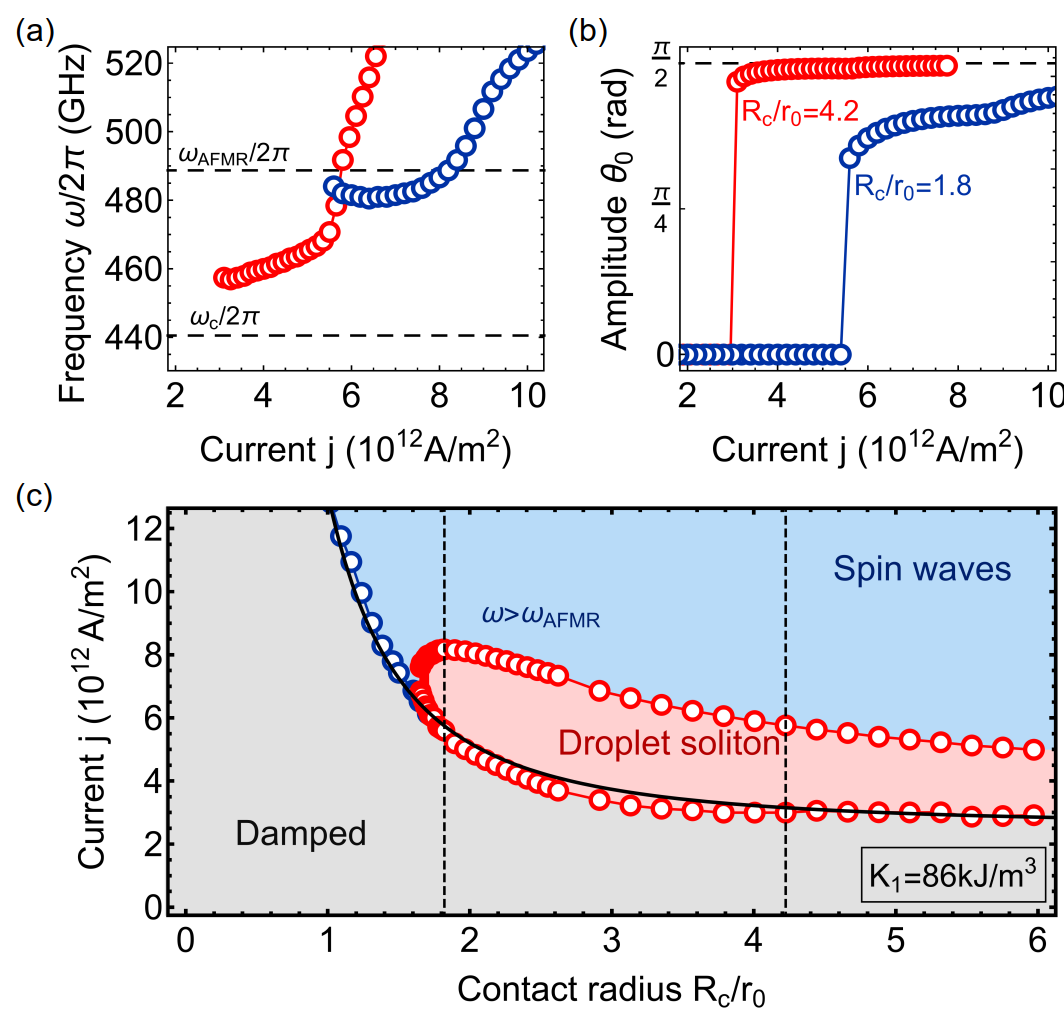}
	\caption{The dependences of (a) the precession frequency $\omega/2\pi$ and (b) the deflection angle $\theta_0$ on the applied current density for the NC with a radius of (red) 73~nm and (blue) 31~nm. (c) The ``phase diagram'' of the excitation type as a function of the NC radius and applied current density; vertical dashed lines indicate selected radii used in (a) and (b). The solid black line is calculated using Eq. (3) from Ref. \onlinecite{Hamdi2022b}. Characteristic length $r_0 = 17.3$~nm.}
	\label{fig:rad_dep}
\end{figure}

In order to excite droplets at smaller NC radii, the droplet's characteristic size should be reduced. To achieve this, we investigated the impact of increasing anisotropy values $K_1$ and $K_2$, keeping their ratio constant at $\rho = K_2/K_1 = 0.3$. Ru and Rh doping would help reduce the droplet characteristic size by increasing the anisotropies and decreasing the exchange field, hence, $c$. The results of simulations, with a NC radius of 43 nm, are depicted in Figure~\ref{fig:anis_results} (c, d). For anisotropy values up to $K_1\approx50$~kJ/m$^3$, the excitation frequency at the threshold exceeds the AFMR, and the localization is not forming. At around $K_1\approx50$ kJ/m$^3$ the localized droplet region is emerging within a narrow range of applied current, but with the subsequent increase in anisotropy, the region of a droplet excitation extends across a broader range of both frequency and current due to the reduction in the characteristic length $r_0$.

Now, we advance to analyze the dependences on NC radius $R_{\text{c}}$. For this, we set the anisotropy constants as $K_1 = 86.0$~kJ/m$^3$ and $K_2 = 25.8$~kJ/m$^3$, thereby determining the characteristic size $r_0$ to equal $17.3$~nm. Figure \ref{fig:rad_dep} shows the diagram of excitation types depending on the $R_{\text{c}}/r_0$ ratio and the applied current density. A scenario solely characterized by the excitation of propagating spin waves is observed for small NC radii, $R_{\text{c}} \lesssim 1.5r_0$. Pure droplets can be excited at NC radii larger than the characteristic size, yet the region is limited by transforming at $j_{\text{sw}}$ to a novel object, where droplet excites propagating spin waves. 
Interestingly, the dependence of the lower threshold on the NC radius corresponds to the one of Slonczewski mode\cite{Slonczewski1999}, which is described in Ref.~\onlinecite{Hamdi2022b} (Eq.3) for the  AFM case. Thus, the maintenance threshold, shown by the solid black line in Fig.~\ref{fig:rad_dep}~(c), is calculated using Eq. (3) in Ref. \onlinecite{Hamdi2022b} with the substitution $\omega_{\text{A}}\rightarrow\omega_{\text{SF}}^2/  \omega_{\text{ex}}$, while the nucleation threshold can be described using $\omega_{\text{A}}\rightarrow \omega_{\text{AFMR}}^2 / \omega_{\text{ex}}$, $\omega_{\text{ex}}=\gamma H_{\text{ex}}$.

The profile of the droplet, as well as the frequency, also depend on the $R_{\text{c}}/r_0$ ratio, and we examine droplet features for two NC radii: $R_{\text{c}} = 1.8r_0$ and $R_{\text{c}} = 4.2r_0$. For a small NC radius, in example $R_{\text{c}}=1.8 r_0$, the frequency range is narrow, akin to the above observations made with smaller anisotropy, see Fig.~\ref{fig:anis_results}. However, the broader current span allows observing a theoretical prediction that suggests a frequency lowering with increased energy influx and, consequently, increased soliton amplitude $\theta_0$. Contrarily, when dealing with a larger NC radius of $R_{\text{c}}=4.2 r_0$, the droplet exhibits a considerably broader frequency range, implying improved tunability. However, this case is characterized by an almost 90$^{\degree}$ precession angle across the current range. This characteristic reduces the output torque $\tau_{\text{out}} = \mathbf{l}\times\dot{\mathbf{l}}$, as the ac component of spin pumping is maximized at $\theta = 45^{\degree}$and diminishes to zero during the proliferation phase\cite{Cheng2016,Hamdi2022b} with 90$^{\degree}$ precession angle. Nevertheless, the total signal could be non-zero since the entire droplet structure under the NC should be considered.

\begin{figure}[hbt!]
\centering
	\includegraphics[width=0.78\linewidth]{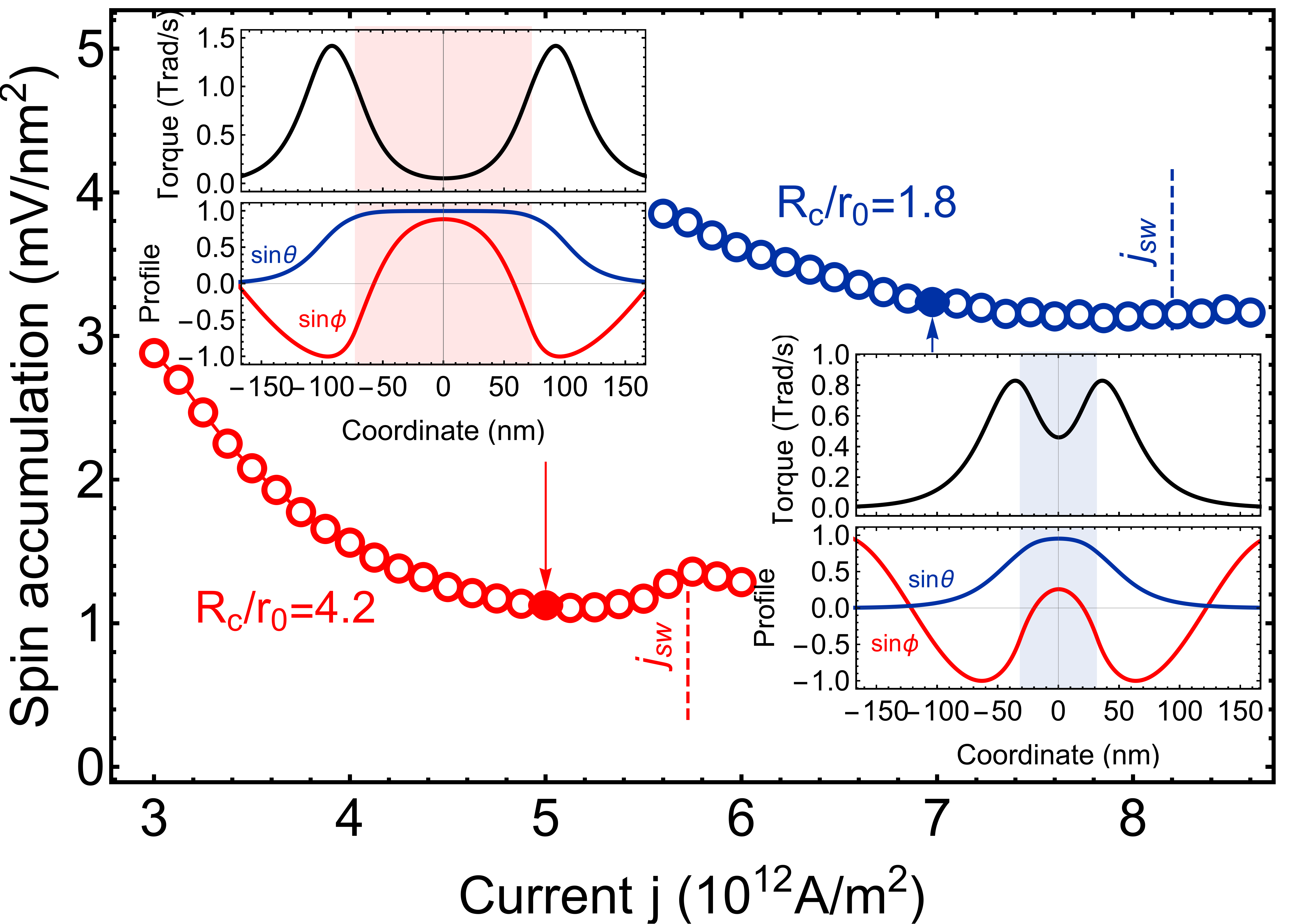}
	\caption{Spin accumulation density, integrated over the NC region, as a function of applied current density for NCs with a radius of (red) 73~nm and (blue) 31~nm. The insets show the cross-section of the output torque distribution and the droplet profile.}
	\label{fig:output}
\end{figure}

For a detailed analysis of the output signals corresponding to chosen NC radii, we computed the dependence of the average spin accumulation on the applied current densities, see Fig.~\ref{fig:output}. The total spin accumulation $V$ is the sum of the output torques generated by the precession of the N\'eel vectors within the NC region $S$, $V=(\hbar / e) \sum_S \tau_{\text{out}}$. We further determined the amplitude of the alternating component using a Fourier transform and normalized it relative to the NC areas to facilitate efficiency comparison. Hence, for a small NC radius $R_{\text{c}}=1.8 r_0$, the absence of planar rotation regions within the droplet profile leads to a higher spin accumulation density, although the NC with $R_{\text{c}}=4.2 r_0$ exhibits a higher total spin accumulation due to its larger interfacial area. 

Note that contrary to the idealized model of droplet in the AFM without dissipation, where in-plane angle $\phi$ is constant in space, our simulations suggest that in the dissipative soliton $\phi$ depends strongly on the radial coordinate $r$. This dependence can be described as the presence of highly nonlinear spin waves (SWs) with a frequency equal to precession frequency $\omega$, confined within the central region of the soliton and propagating from the center of the droplet toward its edge. A qualitative explanation is, in this central area $\theta \simeq \pi/2$ and it can be treated as a local region of a spin-flop state, for which the magnon spectrum has a gapless branch. Thus, for $\omega<\omega_{\text{AFMR}}$ these SWs are localized inside the soliton, and they can propagate outside it only at $\omega>\omega_{\text{AFMR}}$. The wavevector of the SW increases with applied current, which gives rise to the frequency of the droplet (see Fig. \ref{fig:rad_dep}). These SWs can also apply pressure to the transitional areas of the droplet, pushing them beyond the NC region\cite{kim2014propulsion} and reducing the ac output.

\textit{In conclusion}, we demonstrate, both theoretically and through micromagnetic simulations, that AFM dissipative droplet soliton can be excited by applying a spin current through a NC in an extended AFM film featuring first- and second-order uniaxial anisotropy. To stabilize the AFM droplets, the NC radius must be larger than the characteristic size of the soliton.
In this case, for a given anisotropy, the droplet mode is stable above a threshold current and below a switching current, at which point the excited AFM droplet transforms and starts to emit propagating waves. Thus, contrary to ferromagnets, the AFM droplet can be used as an effective emitter of high-frequency magnons propagating with high velocity, which is hard to implement by other methods \cite{hortensius2021coherent}. We show the presence of optimal values of the NC radius for maximizing output spin accumulation density and frequency tunability. We also observe the excitation of nonlinear spin waves inside the droplet, which differs from theoretical predictions under non-dissipative conditions. Based on our results, we suggest Ru and Rh-doped hematite ($\alpha$-Fe$_2$O$_3$) as a perfect material ground for the experimental realization of AFM droplet mode.

\section*{Acknowledgments}

This project is partly funded by the European Research Council (ERC) under the European Union’s Horizon
2020 research and innovation programme (Grant TOPSPIN No 835068) and the Swedish Research Council Framework Grant Dnr. 2016-05980. M.H. thanks Swiss National Science Foundation (SNSF) for financial support via Grant No. 177550.

\section*{Author declarations}

\subsection*{Conflict of Interest}

The authors have no conflicts to disclose.

\section*{Data availability}

The code and output data that support the findings of this study are available from the corresponding author upon reasonable request.

\bibliography{main,library}

\end{document}